\documentclass[preprintnumbers]{revtex4}

\usepackage{graphicx}
\begin{document}
\bibliographystyle{revtex}

\preprint{DFF-377/10/01/
          P3-29}

\title{Degenerate BESS model at  future colliders}

\author{Daniele Dominici}

\email[]{dominici@fi.infn.it}
\affiliation{Dipartimento di Fisica, Universit\`a di Firenze, Sesto F.,
 (Firenze), 50019, Italy}
\affiliation{INFN, Sezione di 
Firenze, Sesto F.,
 (Firenze), 50019, Italy}

\date{\today}

\begin{abstract}
A brief overview of the sensitivity of future colliders
to new vector particles
from strongly interacting Higgs
is presented. In particular the capability
of detecting almost degenerate resonances is reviewed.
\end{abstract}

\maketitle
     
\section{Introduction}
Alternative realizations of the
electroweak symmetry breaking can be formulated by means of effective 
lagrangians which are built on the basis of the
known symmetry properties and which can
a priori contain no resonance or new resonances like
scalar and  vector particles.
The good SM fit to the
electroweak precision data does not necessarily exclude 
possible extensions along this direction which in general assume a large 
Higgs mass.
One can compensate the effect of the large Higgs mass by some new high order
operator or some new particle \cite{vari}. 
A recent critical review of this option  can be found 
in \cite{pesk}. These new operators or the presence of new particles
can give a signature  at new accelerators like LHC and future
linear colliders. For instance 
the parameters $\alpha_4$ and $\alpha_5$ appearing
in the effective lagrangian to order $p^4$ can be detected
by studying $WW$ scattering 
at future colliders with the sensitivity shown 
in \cite{chierici};
the possibility of detecting new vectors from strong $WW$ interaction
at CLIC  \cite{albert} and at VLHC \cite{tao}
has been also recently investigated.
I will present  here a brief overview of
 the phenomenology of new  vector resonances from
the degenerate BESS model.

\section{The degenerate BESS model}
The degenerate BESS model (D-BESS)~\cite{dbess} 
is a realization of dynamical 
electroweak symmetry breaking  which predicts  
the existence of two new 
triplets of gauge bosons almost degenerate in mass ($L^\pm$, $L_3$),
($R^\pm$, $R_3$). The extra parameters  are a new gauge coupling constant
$g''$ and a mass parameter $M$, related to the scale of the
underlying symmetry breaking sector.
In the charged sector the $R^\pm$ fields  are unmixed and $M_{R^\pm}=M$,
while $M_{{L}^\pm}\simeq M (1+x^2)$ where $x=g/g''$ with $g$  the usual 
$SU(2)_W$ gauge coupling constant.
The $L_3$, $R_3$ masses are given by
$M_{L_3}\simeq  M\left(1+ x^2\right),~~ M_{R_3}\simeq
M \left(1+ x^2 \tan^2 \theta\right)$
where $\tan \theta = s_{\theta}/c_{\theta} = g'/g$ and $g'$ is the usual
$U(1)_Y$ gauge coupling constant. These resonances are narrow and almost 
degenerate in mass with $ \Gamma_{L_3}/M\simeq 0.068~ x^2$ and
$\Gamma_{R_3}/M\simeq 0.01~ x^2$, while the neutral mass splitting is:
$\Delta M/M=(M_{L_3}-M_{R_3})/M 
\simeq \left( 1-\tan^2 \theta \right) x^2\simeq 0.70 ~x^2$.
This model respects the existing stringent bounds from electroweak precision 
data since the $S,T,U$ (or $\epsilon_1, \epsilon_2, \epsilon_3$) parameters 
vanish at the leading order due to an additional custodial
symmetry. Therefore the precision electroweak data only set loose bounds on
the parameter space of the model.

Future hadron colliders may be able to discover these new
resonances by their production   through quark annihilation and  decay in
the lepton channel: $q{\bar q'}\to L^\pm,W^\pm\to (e \nu_e)
\mu\nu_\mu$ and $q{\bar q}\to L_3,R_3,Z,\gamma\to 
(e^+e^-)\mu^+\mu^-$.  
The relevant observables are the di-lepton transverse and invariant masses.
The main backgrounds, left to these channels after the lepton isolation
cuts, are the Drell-Yan processes with SM gauge bosons
exchange in the electron and muon channel. A study has been performed 
using Pythia and  CMSJET, which performs a simulation of the energy smearing
of CMS detector,\cite{redi}. Results are given in 
Table~\ref{dom:table1} for the sum of the electron and muon channels
for $L=100$~fb$^{-1}$. For the case $M=$3~TeV the results are given for an 
integrated luminosity of 500 fb$^{-1}$.
\begin{table}
\caption{Sensitivity to $L_3$ and $R_3$ production at the LHC and CLIC 
for $L=$100(500)~fb$^{-1}$ with $M=$1,2(3)~TeV at LHC and
$L=$1000~fb$^{-1}$ at CLIC.}
\begin{tabular}{c c c c c c c}
$g/g''$ & $M$ & $\Gamma_{L_3}$ & $\Gamma_{R_3}$ & $\frac{S}{\sqrt{S+B}}$&
 $S/\sqrt{S+B}$ & $\Delta M$
\\
& (GeV) &(GeV) & (GeV) & LHC ($e+\mu$) & CLIC (hadrons) &  CLIC \\
 \hline  0.1 &
1000 & 0.7 & 0.1 &17.3 & &
\\
0.2 & 1000 & 2.8 & 0.4 & 44.7& &
\\\hline
0.1 & 2000 & 1.4 & 0.2 &3.7& &  
\\
0.2 & 2000 & 5.6 & 0.8 & 8.8& &
 \\\hline
0.1 & 3000 & 2.0 & 0.3 &(3.4)& ~62 & 23.20 $\pm$ .06
\\
0.2 & 3000 & 8.2 & 1.2 &(6.6)& 152 & 83.50 $\pm$ .02
\end{tabular}
\label{dom:table1}
\end{table}

\begin{figure}[htbp]
\centerline{
\includegraphics[scale=0.6]{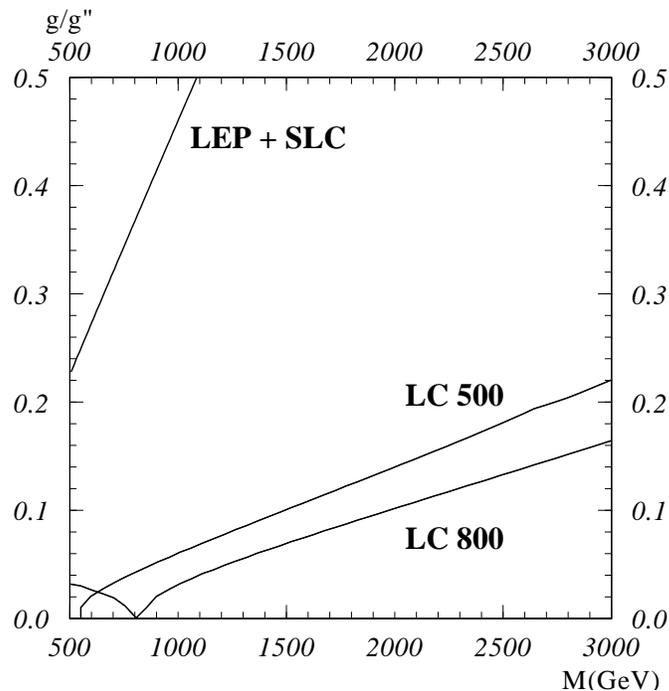}}
\vspace*{0.1cm}
\caption{95\% CL
contour in the plane $(M,~ g/g'')$ from $e^+e^-$ linear
colliders with $\sqrt{s}=500(800)$~GeV. Also shown are the 
present bounds from LEP and SLC. The allowed regions
are below the lines.}
\label{tesla}
\end{figure}

The discovery limit at LHC with $L=100$~fb$^{-1}$ is  $M\sim 2$~TeV
with $g/g''=0.1$. Beyond discovery, the possibility to disentangle the double 
peak structure depends strongly on $g/g''$ and smoothly on the 
mass~\cite{redi}.
A lower energy LC can also probe this multi-TeV region through 
the virtual effects in the cross-sections for $e^+e^-\to {L_3},{R_3},Z,\gamma
\to f \bar f $. Due to the presence of new spin-one resonances the 
annihilation channel in $f \bar f$ and $W^+W^-$ is more efficient than
the fusion channel. In the case of D-BESS, the $L_3$ and $R_3$ states 
are not strongly coupled to $WW$ making the $f\bar f$ final states the most
favourite channel for discovery. The analysis at $\sqrt{s}=$ 500 GeV and 
$\sqrt{s}=$ 800 GeV is based on the following observables:
$\sigma^{\mu},~~\sigma^h$,  $A_{FB}^{e^+e^- \to \mu^+ \mu^-}$,
$A_{FB}^{e^+e^- \to {\bar b} b}$, $A_{LR}^{e^+e^- \to \mu^+
\mu^-}$, $A_{LR}^{e^+e^- \to {\bar b} b}$, $A_{LR}^{e^+e^- \to
{had}}$. For $\sigma^h$ and $\sigma^\mu$  
 a systematic uncertainty of $2\%$ and 1.3$\%$ has been also assumed. 
The sensitivity contours obtained for $L=$ 1000 fb$^{-1}$ and $P(e^-)=80\%$
are shown in Figure~\ref{tesla}. The allowed regions are below the curves.

The LC indirect reach for $\sqrt{s}<$M is 
lower or comparable to that of the LHC. However, the QCD background rejection
essential for the LHC sensitivity still needs to be validated using full 
detector simulation and pile-up effects.

\begin{figure}[htbp]
\centerline{
\includegraphics[scale=0.6]{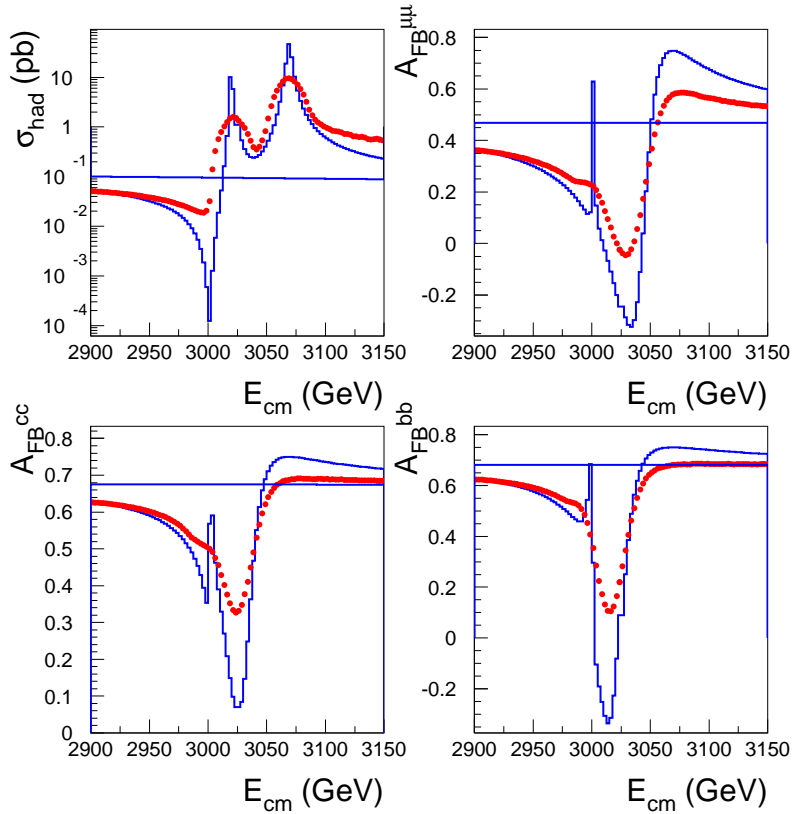}}
\vspace*{0.1cm}
\caption{The hadronic cross section (upper left) and $\mu^+\mu^-$ (upper 
right), $b \bar b$ (lower left) and $c \bar c$ (lower right) forward-backward
asymmetries at energies around 3~TeV. The continous lines represent the 
predictions for the D-BESS model with $M$ = 3~TeV and $g/g''=0.15$, the flat 
lines the SM expectation and the dots the observable D-BESS signal after 
accounting for the CLIC.02 luminosity spectrum}
\label{clic}
\end{figure}

Assuming a resonant signal to be seen at the LHC or at a lower LC, the 
multi-TeV collider can measure its width, mass and investigate the existence 
of an almost degenerate structure~\cite{jhep}. 
This preliminary study has been recently validated
by taking  full account for the luminosity
spectrum and accelerator induced backgrounds\cite{flab}. 
The ability in identifying the 
model distinctive features has been studied using the CLIC
production cross section
and the flavour dependent forward-backward asymmetries, for different values 
of $g/g''$. 
The CLIC 
luminosity spectrum has been obtained with a dedicated beam simulation 
program  for the nominal parameters at $\sqrt{s}$ = 3~TeV. 
In order to study the systematics from the knowledge of this spectrum, 
the modified Yokoya-Chen parameterization  has been adopted. The 
beam energy spectrum is described in terms of $N_\gamma$, the number of 
photons radiated per $e^{\pm}$ in the bunch, the beam energy spread in 
the linac $\sigma_p$ and the fraction $\cal{F}$ of events outside the 5\% of 
the centre-of-mass energy. 
The resulting distributions  for $M$ = 3~TeV and $g/g''=0.15$
are shown in Figure~\ref{clic}
 for the case of 
the CLIC.02 beam parameters
 ($L$=0.40$\times10^{35}$~cm$^{-2}$ s$^{-1}$ and $N_{\gamma}$=1.2).

This study has demonstrated that with 1000~fb$^{-1}$ of data, CLIC will be 
able to resolve the two narrow resonances for values of the coupling ratio 
$g/g'' >$~0.08, corresponding to a mass splitting $\Delta M$ = 13~GeV for 
$M=$ 3~TeV, and to determine $\Delta M$ with a statistical accuracy better 
than 100~MeV (see Table~\ref{dom:table1}).

 The profile of  new resonances can be studied 
with high accuracy due to the large CLIC luminosity.
Additional work has to be done to see how 
this accuracy
can be exploited to distinguish the nature of the resonances.

%
\def\MPL #1 #2 #3 {Mod. Phys. Lett. {\bf#1},\ #2 (#3)}
\def\NPB #1 #2 #3 {Nucl. Phys. {\bf#1},\ #2 (#3)}
\def\PLB #1 #2 #3 {Phys. Lett. {\bf#1},\ #2 (#3)}
\def\PR #1 #2 #3 {Phys. Rep. {\bf#1},\ #2 (#3)}
\def\PRD #1 #2 #3 {Phys. Rev. {\bf#1},\ #2 (#3)}
\def\PRL #1 #2 #3 {Phys. Rev. Lett. {\bf#1},\ #2 (#3)}
\def\RMP #1 #2 #3 {Rev. Mod. Phys. {\bf#1},\ #2 (#3)}
\def\NIM #1 #2 #3 {Nuc. Inst. Meth. {\bf#1},\ #2 (#3)}
\def\ZPC #1 #2 #3 {Z. Phys. {\bf#1},\ #2 (#3)}
\def\EJPC #1 #2 #3 {E. Phys. J. {\bf#1},\ #2 (#3)}
\def\IJMP #1 #2 #3 {Int. J. Mod. Phys. {\bf#1},\ #2 (#3)}
\def\JHEP #1 #2 #3 {J. High En. Phys. {\bf#1},\ #2 (#3)}


\begin{thebibliography}{99}
%
\bibitem{vari}
R.~Casalbuoni, S.~De Curtis, D.~Dominici, R.~Gatto and M.~Grazzini,
\PLB B435 396 1998 ;
R.~Barbieri and A.~Strumia,
\PLB  B462 144 1999 ;
J.~A.~Bagger, A.~F.~Falk and M.~Swartz,
\PRL  84 1385 2000 ;
C.~Kolda and H.~Murayama,
\JHEP 0007 035 2000 ;
R.~S.~Chivukula, C.~Holbling and N.~Evans,
\PRL
 85 511 2000 .
%
\bibitem{pesk}
M.~E.~Peskin and J.~D.~Wells, hep-ph/0101342. 
%
\bibitem{chierici}
R.~Chierici, S.~Rosati and M.~Kobel,
in Physics and Experiments with Future Linear
Colliders, LCWS 2000, Eds. A. Para and H. E. Fisk,
AIP 578, 2001, p. 544;
W.~Kilian,  these proceedings.
%
\bibitem{albert} 
A.~De Roeck, these proceedings.
%
%
\bibitem{tao} 
T.~Han, these proceedings.
%
\bibitem{dbess}  
R.~Casalbuoni {\it et al.}, \PLB B349 533 1995 ,  
\PRD  D53 5201 1996 .
%
\bibitem{redi}
R.~Casalbuoni, S.~De Curtis. and  M.~Redi, \EJPC
 C18 65 2000 .
%
\bibitem{jhep}
R.~Casalbuoni, 
A.~Deandrea, S.~De Curtis, D.~Dominici, R.~Gatto and J.~F.~Gunion,
\JHEP 08  011 1999 .
%
\bibitem{flab}
M.~Battaglia, S.~De Curtis, D.~Dominici, A.~Ferrari and J.~Heikkinen,
in Physics and Experiments with Future Linear
Colliders, LCWS 2000, Eds. A. Para and H. E. Fisk,
AIP 578, 2001, p. 607, hep-ph/0101114.

\end{thebibliography}
\end{document}